\documentclass[12pt]{revtex4}
\usepackage{ulem}
\usepackage{url}
\usepackage{epsfig}
\usepackage{graphicx,color}
\usepackage{epstopdf}
\usepackage[psamsfonts]{amssymb}
\usepackage{amsmath}
\usepackage{indentfirst}

\usepackage{tabularx}

\newcommand{\be}{\begin{equation}}
\newcommand{\ee}{\end{equation}}
\newcommand{\ba}{\begin{eqnarray}}
\newcommand{\ea}{\end{eqnarray}}

\begin{document}

\newcolumntype{L}[1]{>{\raggedright\arraybackslash}p{#1}}
\newcolumntype{C}[1]{>{\centering\arraybackslash}p{#1}}
\newcolumntype{R}[1]{>{\raggedleft\arraybackslash}p{#1}}

\setlength{\tabcolsep}{0.5em} 
{\renewcommand{\arraystretch}{1.2}

\title{{\Large Aristotle vs. Ringelmann}\\Ê{\large On Superlinear Production in Open Source Software}}

\author{Thomas Maillart}
\email{thomas.maillart@unige.ch}
\affiliation{University of Geneva, Switzerland}

\author{Didier Sornette}
\email{dsornette@ethz.ch}
\affiliation{ETH Zurich, Switzerland}

\date{\today}

\begin{abstract}
Organizations exist because they provide additional production gains, in comparison to
horizontal ways of allocating resources, such as markets \cite{coase1937}, and the open source
movement is deemed to be a new kind of {\it peer-production} organization somehow in between
hierarchically organized firms and markets \cite{benkler2002}. However, to strive as {\it a new kind of organization}, open source must provide production gains, which in turn should be measurable. The open source movement is
particularly interesting to study for this reason. Here, we confront and discuss two
contrasting views, which were reported in the literature recently. On the one hand, Sornette et al. 
\cite{sornette2014plosone} uncovered a superlinear production mechanism, which quantifies 
Aristotle adage: {\it ``the whole is more than the sum of its parts"}. On the other hand, Scholtes et al.
\cite{Scholtes2016} found opposite results, and referred to Maximilien Ringelmann, a French
agricultural engineer (1861-1931), who discovered the tendency for individual members of a group to
become increasingly less productive as the size of their group increases \cite{ringelmann1913}. Since 
Ringelmann, the topic of collective intelligence has interested numbers of researchers in social
sciences and social psychology \cite{woolley2010}, as well as practitioners in management aiming at improving the performance of their team \cite{woolley2015}.
In most research and practice case studies, the Ringelmann effect has been found to
hold, while, in contrast, the superlinear effect found by Sornette et al.is novel and may challenge common wisdom  \cite{sornette2014plosone}. 
Here, we compare these two theories, weigh their strengths and weaknesses, and discuss how 
they have been tested with empirical data. We find that they may not contradict each other as much as was claimed by Scholtes et al. \cite{Scholtes2016}.
\end{abstract}

\maketitle

\section{Introduction}
In psychology (Gestalt theory \cite{humphrey1924psychology}), biology (brain functions 
\cite{damoiseaux2009}, ecological networks \cite{jorgensen2012}), physics 
(spontaneous symmetry breaking \cite{anderson1963} and the {\it ``more is different''} concept 
\cite{anderson1972}), and in economics \cite{arthur1994increasing,krugman1996},  the famous adage by 
Aristotle {\it ``the whole is more than the sum of its parts"} has inspired research in complexity science, in
particular regarding emerging behaviors in nature and society \cite{sornette2006critical,perc2013}. 
Indeed, the {\it raison d'\^etre} of societies is 
the prospect that people will achieve more together, yet at some individual 
{\it alienation} costs \cite{locke2014second}. For a society to strive, these alienation costs should overall be smaller 
than the benefits a society can bring to its members. Ideally, fair distribution of benefits should be organized through institutions \cite{ostrom1990} that implement robust mechanisms to enforce cooperation 
\cite{axelrod2006,ostrom1994}. \\

One special instance of a society is the firm. A firm is an organization devoted to production, which 
is born from the internalization of transaction costs associated with gathering production resources, 
in particular human resources: at some point it is less costly to permanently hire an individual whose skills are 
needed often than sourcing them repeatedly on a market \cite{coase1937}. Hence, the employee enters a permanent 
contractual relation and thus, an organizational structure at some alienation costs (less freedom to 
contract with other parties). \\

The open source movement operates in a slightly different fashion:  {\it peer-production} \cite{benkler2002} prescribes that participants to an open source project mainly obey two rules : (i) 
task self-selection and (ii) peer-review. In a nutshell, contribution (i.e., production) enforcement 
mechanisms are very loose, neither relying on hierarchical organization nor market mechanisms and there 
is no clear counterpart to contributions in open source development. The lack of explicit organization
rules in open source has generated much attention in management 
science \cite{lakhani2003,roberts2006}, complex systems and network 
dynamics \cite{maillart2008}, law and economics \cite{sen2005,Lerner2005}, with 
one overarching question being how self-organized communities gather and, moreover, 
produce efficiently together in absence of organizational rules clearly tied to incentives \cite{vonkrogh2012}.\\

The open source movement gathers people with heterogeneous incentives, ranging from
hedonism to paid jobs \cite{vonkrogh2012}. It is therefore difficult to measure the implications of individual, and of collective intelligence and coordination, on the production of 
source code. In particular, there is the question of how cumulative innovation emerges from self-selected 
contributions and peer-review, which on average make software more robust and help the emergence 
of new functionalities. Measuring production and productivity of collective intelligence may be a significant addition to the
debate, and attempts to measure productivity of software developers is nearly as old as the software
industry \cite{brooks1975}, with several models developed to measure the efficiency of software programmers,
yet with the assumption that programmers work in a corporate environment, which is usually highly 
scheduled. \\

The bottom-up and collective intelligence aspects of production have been much less covered, in open 
source and more generally, in open collaboration \cite{chesbrough2006}. Dealing with groups such as firms and production 
units, management science also aims to understand when and how a group 
can produce more than the sum of its individual contributions, and to design ways to improve team performance
\cite{tziner1985,sundstrom1990work,cohen1997,neuman1999team}, 
through the mechanism of complementarity in organization
\cite{ennen2010,lin2006communities} and innovations \cite{sacramento2006team}. 
Because most activities in our modern environment require coordination and
collaborative actions within groups of widely varying sizes, 
it is the fundamental aspiration of any manager, be it in the public or private sector, to find and master the 
determinants of enhanced productivity.  Since  Ringelmann, the topic of collective intelligence has interested numbers of researchers in social sciences and social psychology \cite{woolley2010}, as well as practitioners in management aiming at improving the performance of their team \cite{woolley2015}.

Despite their conflicting views, the contributions by Sornette et al. 
\cite{sornette2014plosone} and Scholtes 
et al. \cite{Scholtes2016} provide key insights on that matter, in particular, yet not limited to, open 
source development. We focus on these two papers in particular because Scholtes 
et al. \cite{Scholtes2016} challenged evidence brought forth by Sornette et al. 
\cite{sornette2014plosone}, creating confusion or perhaps even worse the sentiment that
the superlinear productivity law is stillborn, having been killed just out of its academic womb.
In the remaining of this paper, we describe and compare the two approaches (Section 
\ref{sec:production_measures}), then discuss the strengths and weaknesses of each 
approach (Section \ref{sec:discussion}), and conclude (Section \ref{sec:conclusion}).

\section{Production and Productivity Measures for Open Source Software}
\label{sec:production_measures}

Here, we expose the two contrasting perspectives taken by Scholtes 
et al. \cite{Scholtes2016} on the one hand, and by Sornette et al. 
\cite{sornette2014plosone} on the other hand.

\subsection{The Ringelmann effect in software engineering}

There is a common wisdom supported by a vast majority of studies, which tend to show that teams of software developers become
less productive as they get bigger. In empirical software engineering, this phenomenon is known as the {\it BrooksÕ law of software 
project management}, which states that {\it ``adding manpower to a late software project makes it later"}  \cite{brooks1975}. The identified cause for the Brooks' law 
is the increasing coordination costs involved as teams get larger. In social psychology, this phenomenon is also known as the
Ringelmann effect, in reference to Maximilien Ringelmann, a French agricultural engineer (1861-1931) who discovered the tendency for
individual members of a group to become increasingly less productive as the size of their group increases \cite{ringelmann1913}.\\

Scholtes et al. \cite{Scholtes2016} performed a study using a dataset of 58 open source software (OSS) projects, which amount in 
total to more than 580,000 commits contributed by more than 30,000 developers. Their study indeed finds that the Ringelmann effect seems to 
hold on {\bf average}. Here is the way they proceeded for their study. While in structured organizations, a {\it team } can be easily defined and 
measured, a team in OSS projects is more complicated. Indeed, Scholtes et al.  \cite{Scholtes2016} reported that 40\% of contributions to 
OSS projects (i.e., {\it commits}) were made by one-time contributors. Researchers have identified different circles of contributors 
from a core team (producing up to 90\% of the source code), to less involved contributors, to one-shot 
contributors, and finally, to lurkers, who follow the advancement of a project without contributing to the source code, and yet participating e.g., 
on the mailing list or posting issues \cite{david2008}. The heterogeneous, distributed, and uneven proportion of contributions makes the study of 
OSS project organizations complicated, particularly across projects, themselves of heterogeneous nature. \\

Scholtes et al. provided a dynamic formulation of what a team is, considering that a developer, who has not contributed after 295 days has 10\% 
chance to further contribute. Therefore, they chose a window of 295 days to define a team size at time $t$, which is the count of 
contributors who have committed at least once in the last 295 days. The output must also be measured. Various measures of source code 
production have been developed to account for contribution effort \cite{boehm1984,boehm2000}. Scholtes et al. decided to focus on
quantifying changes, as measured by the edit 
distance (also called the Levenshtein distance \cite{levenshtein1966}), i.e.,
the minimum number of bytes one has to permute/add/delete to compare between a version of the source code and a committed update. 
Scholtes et al. used averaged contributions over time windows of 7 days (rationalized by the fact that in 90\% of the cases, two consecutive 
commits occur within this time window).\\

\begin{table}
\begin{center}
\begin{tabular}{|L{5cm}|C{5cm}|C{5cm}|}
\hline
  &~~ {\bf Scholtes et al.} ~&~~{\bf Sornette et al.} ~~\\
  \hline
{\bf OSS Projects studied} & 56 & 164 \\
  \hline
  {\bf Number of Contributors} & $30,845$& $15,650$ \\
\hline
  {\bf Number of Commits} & $581,353$ & $8,220,673$\\
  \hline
 {\bf Project sampling} &  
 \begin{itemize}\setlength\itemsep{-0.5em}\vspace{-0.7cm}
 \item $>$ 1 year of activity
 \item $>$ 50 active developers 
 \item $\subset$ 100 most popular projects in GitHub
 \end{itemize} 
 & \multicolumn{1}{|L{5cm}|}{ Random sampling : Power law distribution found $Pr(size > S) \sim 1/S^{1.4}$ with $S$ the number of contributors} \\
  \hline
\end{tabular}
\caption{Comparison of datasets used by Scholtes et al, and Sornette et al.}
\end{center}
\end{table}

Scholtes et al. first measured the output (i.e., number of commits and contributions as defined by edit 
distance over the last 7 days) as a function of the input (i.e., active developers within the same 7 days window at time $t$). They found that, when
the number of developers increase, the mean 
contribution per developer decreases. Moreover, when considering mean contribution per active developer as a function of team size 
(i.e., developers active in the last 295 days), the results 
show the same negative scaling function [for commits : $\sim team^{-0.24}$ ($p<0.001$ and $r^2 = 0.16$) and for contributions $\sim team^{-0.36}$ 
($p<0.001$ and $r^2 = 0.08$), obtained by MM-estimation]. The authors highlight the particularly low $r^2$ in both case, 
reflecting the high variability of their average measures, and as such, they conclude that it is impossible to make robust predictions from 
these scaling laws. 
Considering the output as a function of team size (here, the input is considered as the amount of resources available, i.e., contributors who 
have roughly more than a 10\% chance to contribute), again negative scaling properties are found [for commits : $~team^{-0.75}$
($p<0.001$ and $r^2 = 0.44$) and for contributions :$~team^{-0.86}$ ($p<0.001$ and $r^2 = 0.25$), obtained by MM-estimation].\\

As a consequence, Scholtes et al. concluded that {\it ``OSS communities are indeed no magical exception from the basic economics of collaborative 
software engineering"}, and they further attempted to substantiate the observed decreasing return on scale, considering two commonly 
accepted reasons for the Ringelmann effect : (i) free-loading and (ii) coordination costs. They concentrated on the latter because there is a 
substantial body of evidence and research work on coordination in software engineering. Although the authors did not mention it, it is indeed 
hard, if not impossible, to define free-loading when contributors are actually not compelled to contribute (following the general rules of peer-production applicable in OSS).
To assess coordination effort and its effects on productivity, Scholtes et al. borrowed from Cataldo et al. \cite{cataldo2006} and computed the 
co-edition directed network for all developers (direction stands for chronological influence) as a function of time (i.e., time windows of 7 days), 
with a distinction between out-degree $k_{out}$  (i.e., one developer has to build on changes by $k_{out}$ other developers) and in-degree 
$k_{in}$ (i.e., $k_{in}$ developers must build on changes by one developer). Scholtes et al. considered first the mean out-degree as a function  
of the size of the coordination network \footnote{the size of the coordination is not clearly defined in the paper. We assume that the authors 
talked about the number of network edges in each time window of 7 days}. The  mean out-degree and the size of the coordination network 
seem to be positively correlated, but it is not clear what we can learn from this result (Figure 11 in Scholtes et al. \cite{Scholtes2016}; not 
discussed in the paper). Finally, Scholtes et al. considered mean out-degree as a function of the negative productivity scaling exponent 
described above. They found that projects with {\it ``strongly negative and significant slopes for the scaling of productivity also exhibit 
pronouncedly positive scaling exponents for the growth of the mean (weighted) out-degree"} (Figure 12 in Scholtes et al. \cite{Scholtes2016}).\\

Based on these results, Scholtes et al. asserted that OSS projects exhibit dis-economies of scale in production as a function of team size, and 
hence, sub-linear productivity. They rejected the evidence that {\it ``the whole is more than the sum of its parts''} 
evidenced by the superlinear productivity shown by Sornette et al. \cite{sornette2014plosone}.

\begin{table}
\begin{center}
\begin{tabular}{|L{4cm}|C{6cm}|C{5cm}|}
\hline
  &~~ {\bf Scholtes et al.} ~&~~{\bf Sornette et al.} ~~\\
  \hline
 {\bf Activity window} & 7 days & 5 days\\
\hline
{\bf Team definition} & $s :=$ at least one contribution in last 295 days & all developers \\
  \hline
{\bf Active developers } & within 7 days windows & $c$ (within 5 days windows) \\
  \hline
  {\bf Primary production measure} & commit contributions (Levenshtein distance between commits) & $R :=$ number of commits  \\
  \hline
  {\bf Productivity} & $n:=$ mean number of commits;  $c:=$ mean number of commit contributions per
active developer & number of commits $R$ performed  by $c$ active developers \\
  \hline
  {\bf Productivity scaling properties} & 
  $\begin{aligned}[t] 
n \sim s^{\alpha_0}~with~\alpha_0 \approx -0.24\\
c \sim s^{\alpha_1}~with~\alpha_1 \approx -0.36\\
\end{aligned}$
   & $ R \sim c^{\beta}~with~\hat{\beta} \approx 4/3$\\
  \hline
 \end{tabular}
\caption{Comparison of definitions and results by Scholtes et al., and Sornette et al. Comparing both approach, we observe that resorting to averaging (Scholtes et al.) leads to tremendously different results, compared to a direct measure of production per developer (Sornette et al.).}
\end{center}
\end{table}

\subsection{The Aristotle effect perspective}
In contrast to the previous section, by analyzing 164 open source software (OSS) projects of broadly 
distributed sizes ranging from 5 to 1,678 contributors, Sornette et al. 
\cite{sornette2014plosone} found that contribution activity $R$, defined in terms of number of commits, within a time window 
of 5 days, is a superlinear function $R \sim c^{\beta}$ of the number of active developers $c$ during the same period. 
The superlinear exponent is on average $\hat{\beta} \approx 
4/3$, over all projects studied, with a rather large variability with $\hat{\beta}$ ranging from $1$ to $3$. 
They found that $\hat{\beta}$ tends to decrease 
with the number of contributors in the five day window, fluctuating around $1$ or less for 
more than 30 to 50 contributors. Moreover, as reported in Sornette et al. \cite{sornette2014plosone}, the distribution of 
total number of developers per project is heavy-tailed, 
i.e., with many small projects and a few very large ones.\\

Sornette et al. explored two possible mechanisms generating the observed superlinear
phenomenon : (i) an interaction-based mechanism (including interactions leading to a phase transition or 
to a super-radiance phenomenon \cite{GrossHaroche82}) and (ii) a large-deviation mechanism, 
based on the fact that, in the presence of a heavy-tailed distribution of contributors per project, 
many developers contribute just few commits while a minority contribute most of the 
commits; then, the larger the group size, the more likely it is for a large contributor to be present, 
leading to the superlinearity phenomenon. The observation that a few developers dominate the
overall contribution is well-known in OSS, and is also reported by Scholtes et al. Sornette et al. did not attempt to 
distinguish which one of these two mechanisms might be at work.
They however considered that both the interaction-based and the large deviations mechanisms can be captured  together by a generic cascade process, which has been found to be well described by self-excited Hawkes conditional Poisson processes
\cite{hawkes1974acluster}, in particular for human dynamics \cite{mohler2011,baldwin2012,ait-sahalia2010,filiminov2012,filiminov2014}, taking into 
account the specifics of human timing \cite{maillart2011}. \\

The Hawkes process is defined by the conditional point process intensity $I(t)$ of events (commits) given by
\be
I(t)= \lambda(t) + \sum_{i | t_t<t}  f_i \phi(t-t_i)~,
\label{jruym}
\ee
where $\{t_i, i=1, 2, ...\}$ are the timestamps of past commits, $\lambda(t)$ is the spontaneous
exogenous rate of commits, $f_{i}$ is the 
fertility of commit $i$ that quantifies the number of commits (of first generation)
that it can potentially trigger directly, and $ \phi(t-t_i)$ is the memory kernel, whose
integral is normalized to $1$, which weights how 
much past commit activities influence future ones. 
According to (\ref{jruym}), the number of commits contributed between time $t$ and $t+dt$
results from two sources: (i) an exogenous source $\lambda(t) dt$ representing the spontaneous commits not related
to previous commits; (ii) an endogenous term represented by the sum over all commits that were 
made prior to $t$, and which are susceptible to trigger future commits. The class of Hawkes models can be mapped onto the general class of branching processes \cite{daley2007}. 
The statistical average fertility $\langle f_i \rangle$ defines the branching ratio $n$, which is the key
parameter. For $n<1$, $n=1$ and $n>1$, the process is respectively sub-critical, critical and super-critical 
\cite{helmstetter2002subcritical,helmstetter2003}. Interpreting a cluster or connected cascade in a given branching process 
of triggered contributions as the burst of production in a group of developers, the distribution of contributions is thus mapped onto that of triggered cluster sizes \cite{saichev2005power}. \\

Sornette et al. found and empirically validated that, at criticality, there is a relationship between the power 
law tail distribution (with exponent $\gamma 
\approx 1.5$) of activity per contributors per time bin of 5 days, the power law tail 
distribution of cluster size, which is equivalent to the production $R$ per 
contributor with renormalized exponent $\mu = 1/\gamma$ and the superlinear scaling 
exponent $\beta = \gamma = 1/\gamma$. However, as already mentioned, Sornette et al. 
found that the superlinear scaling exponent $\beta$ tends to decrease as a function of the 
total number of contributors in an OSS project. Likewise, the 
frequency of productive bursts is reduced for larger projects, suggesting that large projects bear additional coordination costs.

\section{Aristotle versus Ringelmann ?}
\label{sec:discussion}
Before considering the fundamental differences between the two approaches presented here, and their validity, we shall highlight some results, which to some extent bear 
resemblance. Sornette et al. found that large projects tend to exhibit less powerful and less frequent superlinear productive bursts. 
This result may look similar to the findings by Scholtes et al., who studied only large projects. However, Sornette et al. do not say that there are dis-economies of scale, but 
rather that economies of scales appear to be weaker. Similarly to the {\it co-edition directed network} model developed by Scholtes et al., the self-excited 
Hawkes conditional Poisson process measures how past commits influence future commits. It does so in a way that incorporates the influence all past events, while the network approach by Scholtes et al. relies only on 7-day 
contribution windows. In other words, Scholtes et al. considered that, in order to be dependent (and bear coordination costs between each other), two commits must occur within the same time window. The network approach, taking into account who 
changed which file, brings more information regarding how contributions relate to each other. 
It is also interesting to note the closeness of the short- and 
long-term time windows used in the two studies : 7 and 295 days for Scholtes et al. versus 5 and 250 days for Sornette et al. 
While Scholtes et al.
provide a justification, Sornette et al. are only concerned with robustness and check that the same results are obtained by varying the short-term window. In Sornette et al., no rationale  is provided for the long-term window.\\

Beyond these resemblance and arguably a common research question, nearly every other aspects differ in the two 
studies: the chosen approach, the definitions of productivity, and the data used. 
This raises a number of questions on the main claim by Scholtes et al. 
that Sornette et al. were wrong about the superlinear productive bursts \cite{Scholtes2016}. 
In the following, we thus highlight and discuss these methodological divergences. 
Moreover, we discuss the approach chosen by Scholtes et al. to take on the results by 
Sornette et al, in an era that promotes open science and, most importantly, reproducibility of scientific results.

\subsection{Productivity \& Team Size}
We start by a fundamental conceptual remark that illuminates one key difference between the approach of
Scholtes et al. \cite{Scholtes2016} and the one by Sornette et al. \cite{sornette2014plosone}. Scholtes et al. consider production
in the mean, using as metric the {\it average output per team member} (Introduction, 2nd paragraph, line 3), and argue
that it increases when synergy effects are present and decrease due to 
{\it communication and coordination overhead} (which surges with larger teams).
In contrast, Sornette et al. argued and demonstrated that using an average output is 
misleading in the presence of highly bursty dynamics characterized by power law tail distributions
with small tail exponents. This empirical fact is also cited by Scholtes et al. and
well-documented in the open source software production literature
and for other open collaboration projects, such as wikis and Wikipedia \cite{robles2004,hindle2008,alali2008,arafat2009}.
In open collaboration, a few contributors account for a majority of performed work, whether counted in lines of code, commits, files modified, and so on. This is one of the features associated with the fact that the distribution of contributions, 
counted in commits or in lines of codes, possesses a power law tail of the form $P(X>x) \sim 1/x^{\mu}$ with $\mu < 1$ \cite{sornette2014plosone}. Such distributions are {\it wild}  \cite{mandelbrot2007} in the sense that 
their two first statistical moments (mean and variance) are undefined and diverge as the sample grows. 
For such heavy-tail distributions, reasoning in mean is fundamentally erroneous, as Scholtes et al. could indeed experience when trying to perform predictions (c.f., Section 4.1 in 
\cite{Scholtes2016}). For a finite number $n$ of developers in the project, it is easy to show that the average production scales as $\sim n^{1/\mu}$ for $
\mu < 1$ and $\sim n$ for $\mu \geq 1$.
Defining productivity as the ratio of the total production by the number $n$ of team members,
this shows that productivity scales as $\sim \frac{n^{1/\mu}}{n} = n^{{1 \over \mu} - 1}$ for $\mu < 1$ 
and is constant for $\mu \geq 1$. This latter case is the null hypothesis of an approximate 
constant output per team member. Superlinear production is quantified by $\mu < 1$, leading to 
a growing productivity per team member, the larger the team. 
Searching for a superlinear productivity is different from seeking a superlinear production, the former
requiring ${1 \over \mu} - 1 >1$, i.e., $\mu < 1/2$, while the later just needs $\mu < 1$. 
In their dataset of 164 projects, Sornette et al. found that only four projects are characterized by $\mu < 0.5$ (while most
obey $\mu<1$). 
Scholtes et al. have used a much smaller dataset of 58 projects, which implies a $\approx 0.8\%$ chance to find one project for which the average productivity scales superlinearly with team size. \\

More generally, the definition of productivity needs to be carefully addressed.
Indeed, an open source software community does not come into being fully grown. It starts rather small 
and then grows progressively -- one could say {\it organically} -- with the project. When growing, the community bears increasing communication and coordination costs as pointed out by Scholtes et al. While recognizing the importance of different team sizes, Scholtes et al., picked projects meeting the following criteria: (i) at least one year of activity, (ii) 50 different active developers, and (iii) being among the 100 most popular projects, as measured by the number of forks on GitHub, a leading online service for open source software production. In contrast, Sornette et al. chose a representative sample of the open source ecosystem with 134 projects with less than 50 developers and 30 projects with more than 50 developers (with a minimum of 5 developers). The representative sampling of projects (see Figure 1 in \cite{sornette2014plosone}) showed that the superlinear production 
is usually valid only for projects of sizes no more than 30 to 50 members who are active at a given time. Sornette et al. found statistically significant
evidence that the superlinear production tends to fade away to just linear production (i.e., constant productivity
per developer) for projects with more than 50 developers (see Figure 8 in \cite{sornette2014plosone}). 
In other words, the sample selection made by Scholtes et al. seems heavily biased towards large projects, which represent the few large (presumably older) projects and are indeed exposed to more communication and coordination costs, and also exhibit less synergy effects. \\

More specifically, Scholtes et al. defined a team as the set of developers who are active at least once
within a time window of 295 days, determined by the 90th quantile of the distribution of times between 
two consecutive commits by the same person. This definition excludes
developers with a unique contribution, who  nevertheless account for 40\% of all contributions, as 
reported by Scholtes et al. in Section 3.2 of \cite{Scholtes2016} (end of second paragraph). 
In line with our above remarks concerning the heavy-tailed distribution of contribution sizes, 
this definition amounts to {\it throw away the baby with the bath}, since it is fundamentally
ill-suited to account for the fact that a few, often most senior, developers  
may not contribute for years in between two commits (see Figure 2 in \cite{saichev2013}), while
at the same time they may account for most of the contribution production.
The definition of contributors proposed by Scholtes et al. is thus biased with respect to the special nature of the open source software community, which 
is, almost by essence, different from a corporate organization, as documented in a number of management science articles (see e.g., 
\cite{vonkrogh2012} and references therein).\\

Yet, productivity may be defined in a variety of ways, each with their advantages and shortcomings.
Scholtes et al. considered productivity as production per active developers among a team (defined as an aggregate of working developers in large -- 295 days -- time windows), while 
Sornette et al. considered productivity as production per developer and per time unit (i.e., over a short time period of 5 days). Even though not perfect, the 
latter definition is more fine-grained than the one proposed by Scholtes et al., and precisely allows capturing the subtle highly non-linear bursts of activity 
reported in \cite{sornette2014plosone}, which could not be observed by averaging developer engagement (over a team aggregate and over time). In 
essence, two visions oppose each other: Scholtes et al. adopted a {\it software engineering} perspective, which takes roots in the necessity to 
measure the effort and productivity by software developers in a competitive industry. Because of the complexity of information systems, and the 
importance of outsourcing, the software development industry may suffer from the principal-agent problem \cite{keil2005} and hence, may require controlling.  
The {\it software engineering} perspective is reflected by the sampling of only large projects (suggesting that smaller projects are not really worth 
studying), and by the definition of active contributors and teams, which ignores 40\% of the contributors. On the contrary, Sornette et al. considered the 
OSS ecosystem with no filter, taking a more general approach in project sampling, in the definition of contributions, and in theory elaboration and 
validation.\\

\subsection{Commits \& superlinear scaling}

Productivity is the ratio of an output and an input. So far, we have mainly discussed the developer input, 
i.e., the human capital. Scholtes et al. raised concerns about the output, and claimed that the number of commits is an {\it erroneous} measure of production. 
For that, they bring forth the following argument: {\it the total number of commits contributed by $n$ developers active in a given time period cannot -- by definition -- 
be less than $n$, which is why the total number of commits must scale {\it at least} linearly with team size.} This apparently common-sensical claim is incorrect as we demonstrate here.
Let us consider $n$ developers. The largest contributor makes $N$ commits (resp. lines of code). The second one contributes $N/2^{\alpha}$ commits. The third one contributes $N/3^{\alpha}$ 
commits, and the $n$-th one contributes $N/n^{\alpha}$ commits. If $0 < \alpha < 1$ and $n$ and $N$ are such that $N/n^{\alpha} \geqslant 1$ (i.e., $n \leqslant N^{1/\alpha}$), then the total number 
of commits contributed by $n$ developers is given by 
\begin{equation}
S(n)  =  \frac{N}{1^{\alpha}} +  \frac{N}{2^{\alpha}} +  \frac{N}{3^{\alpha}} + \cdots+  \frac{N}{j^{\alpha}} + \cdots + \frac{N}{n^{\alpha}}    \sim   N \cdot n^{1-\alpha}.
\end{equation}
Thus, in this example, the total contributions of these developers grow sub-linearly as a function of group size $n$, with exponent $1-\alpha$. 
Let us illustrate this demonstration by a numerical example, showing that the sublinear effect
is clearly visible even for small team sizes. Let us assume that $N=10$ and $\alpha =1/2$. For $n=5$ developers, the total number of commits is $32$. For $n=25$, the total  contribution is equal to 
$86$ commits, which is 2.7 times that for the team of 5 developers (and not 5 times more). Note that for the team of $25$ developers, the first contributor makes 10 commits and the last one 
contributes $2$ commits. We believe Scholtes et al. made a very common confusion between absolute numbers and scaling properties. More generally, in the field of fractals, this error is also often 
found in the literature that confuses the fact that the fractal dimension (here, the scaling exponent) tells nothing (or very little) about the density (here, the number of commits per developer).\\

Dismissing commits as a measure of production, Scholtes et al. used the Levenshtein edit distance \cite{levenshtein1966} of source code changes between two consecutive commits (i.e., so-called {\it diffs}). The Levenshtein edit distance counts the number of permutations, additions and deletions of characters necessary to match two different strings. Using the Levenshtein edit distance is without doubt more detailed, but it is not sufficient to dismiss commits. Even though they have not used the Levenshtein edit distance, Sornette et al. showed that superlinear scaling production holds as well for lines of code (see Figure 3 in \cite{sornette2014plosone}), up to an additional scaling factor that defines the relation between commits and lines of code. In order to properly dismiss commits as a measure of contribution, 
Scholtes et al. may have wanted to show that there is no relation between commits and their contribution metric, for which there is no clear consensus in the scientific literature. The Levenshtein distance is more detailed than commits, but may not necessarily contain additional relevant information. Moreover, at a qualitative level, we should stress that using the more detailed Levenshtein edit distance is not without its own problems.  One may indeed argue that changing one character or a single line of code in a piece of software, while quantified as minor by 
the Levenshtein distance, could be in some cases 
a tremendous output reflecting a major commitment in terms of human capital (think e.g., of a small edit correcting a security vulnerability) \cite{kuypers2016,maillart2017,kuypers2018}. We suggest that a truly faithful measure of input would be the time effectively spent in front of a computer by a contributor in order to achieve a task for the focal open source software project. Unfortunately, this information is not available to open source software researchers and, even if it would be available, one could endlessly debate on a broad (resp. narrow) definition of time consumption, and whether the coffee break and the ping-pong sessions are actually parts of the production time: nearly all Silicon Valley software companies would include this time as truly productive time. Another way of proceeding would be to use a robust approach to attribute value to each contribution instead of assuming value. Such an approach to attribute value to contributions has been previously proposed by Maillart et Sornette \cite{maillart2014}. In more conditioned environments, other ways to attribute contribution value to individuals engaged in collective intelligence have been tested and studied \cite{gulley2010}.\\
 
\subsection{From Aristotle to Ringelmann : a missed opportunity for reproducible science ?}
Open science is nowadays highly promoted to ensure reproducibility of scientific results, and to encourage research groups to {\it ``debug"} and build upon 
each other works \cite{nosek2015,munafo2017}. The open science movement is inspired by the open source software movement, best summarized by 
the seminal adage : {\it Given enough eyeballs, all bugs are shallow} \cite{raymond1999}. The authors of the paper {\it From Aristotle to Ringelmann: a large 
scale analysis of team productivity and coordination in Open Source Software projects} are (or were) members of the Chair of Systems Design. The Chair 
of Systems Design has been known to be a pioneering research group at ETH Zurich, advocating the use of open source software and contributing 
significantly to the open access movement. Sornette et al. published in PlosOne, which the first and leading open access scientific journal. Along with the paper, they submitted 
and shared the data they used for their study. Scholtes et al. clearly framed their paper as a response to Sornette et al. ({\it ``From Aristotle to Ringelmann"}
in the title), and they claimed that the results published by Sornette et al. do not hold. The claim by Scholtes et al. is at best misleading as they did not 
bind to elementary principles of science reproducibility. First, they neither used the same data nor detailed 
the potential implications of using a different dataset. Second, they neither invalidated the method by Sornette et al. nor compared thoroughly both 
approaches, with their pros and cons, as we have done above. Third and foremost, 
they did not bring compelling arguments for changing the assumptions
underlying the analyses. These limitations 
deeply undermine their claims that the results of Sornette et al. are incorrect, as we have shown above. 
As a result, it is challenging to weigh the value of one approach against the other, 
and in this regard, limits the pertinence of the contribution by Scholtes et al.\\

 Scholtes et al. submitted to and published 
their paper in the {\it Journal Empirical Software Engineering}. First, one would expect that claims questioning the 
validity of the results obtained by Sornette et al.  should have been sent to the same journal (i.e., PlosOne), as a comment 
or a follow-up paper to the editors. Second, it is rather surprising that the editors and 
the reviewers of the {\it Journal Empirical Software Engineering} did not raise any issue concerning the approach by Scholtes et al. to rebut the findings by Sornette et al., in particular given the many problems that we have highlighted above.
Third, when the present authors attempted to send an earlier version of this manuscript \cite{maillart2016} as a response to the editor, they received the following response from  the editors of the {\it Journal Empirical Software Engineering}:  {\it ``the [...] journal does not publish any responses to articles. We encourage you to expand the response to a full research paper, e.g., by repeating the experiments, adding additional research questions, etc"}. In other words, the editors barred the possibility to react to Scholtes et al. in their journal, and they asked the present authors to perform what they should have requested at first for the manuscript by Scholtes et al. 

\section{Conclusion}
\label{sec:conclusion}
We have carefully described the two methods and results by Sornette et al. \cite{sornette2014plosone} and Scholtes et al. \cite{Scholtes2016},  and 
their apparent opposite results  (i.e., the Aristotle vs. Ringelmann effects), with emphasis on their commonalities and differences. Despite claiming that 
the results by Sornette et al. do not hold, Scholtes et al did not use the same data (made available by Sornette et al. following open access standards), they used a 
totally different methodology (based on averages) that does not allow a direct testing of the methods and results by Sornette et al. 
(designed to be able to quantify bursty dynamics and large deviations). 
However compelling and probably valid in its own way, the method followed by Scholtes et al. does 
not help directly invalidate the results and theory by 
Sornette et al. We believe there is much room for the {\it Aristotle vs. Ringelmann} debate, and we are glad that Scholtes et al. took upon the challenge. Our conclusion is that Sornette et al.'s results hold for no more than 30-50 contributors working simultaneously, while
Scholtes et al.'s results may apply for larger projects.
Yet, we believe that proceeding in a way that follows good practices regarding open and reproducible science, as well as using a more standard 
publication channel for their challenge, would have helped developing a much more data grounded, constructive and serene debate.

\section*{Acknowledgements}
Thomas Maillart acknowledges support from the Swiss National Science Foundation (grants P3P3P2\_167694 and P300P2\_158462).

\bibliographystyle{apsrev4-1}

\end{document}